\documentclass[3p,times,twocolumn]{elsarticle}
\usepackage{ecrc}
\volume{00}
\firstpage{1}
\journalname{Nuclear and Particle Physics Proceedings}
\runauth{G. Vujanovic et al.}
\jid{nppp}
\jnltitlelogo{Nuclear and Particle Physics Proceedings}
\usepackage{amssymb}
\usepackage[figuresright]{rotating}
\begin{document}
\begin{frontmatter}
\title{Dilepton radiation and bulk viscosity in heavy-ion collisions}
\author[label1]{Gojko Vujanovic}
\address[label1]{Department of Physics, The Ohio State University, 191 West Woodruff Avenue, Columbus, Ohio 43210, USA}
\author[label2]{Jean-Fran\c cois Paquet}
\address[label2]{Department of Physics \& Astronomy, Stony Brook University, Stony Brook, New York 11794, USA }
\author[label3,label4]{Chun Shen}
\address[label3]{Department of Physics, McGill University, 3600 University Street, Montr\'eal, QC, H3A 2T8, Canada}
\address[label4]{Department of Physics, Brookhaven National Laboratory, Upton, New York
11973-5000}
\author[label5]{Gabriel S. Denicol}
\address[label5]{Instituto de F\'{i}sica, Universidade Federal Fluminense, UFF, Niter\'{o}i, 24210-346, RJ, Brazil}
\author[label3]{Sangyong Jeon}
\author[label3]{Charles Gale}
\author[label1]{Ulrich Heinz}

\begin{abstract}
Starting from IP-Glasma initial conditions, we investigate the effects of bulk pressure on thermal dilepton production at the Relativistic Heavy Ion Collider (RHIC) and the Large Hadron Collider (LHC) energies. Though results of the thermal dilepton $v_2$ under the influence of both bulk and shear viscosity is presented for top RHIC energy, more emphasis is put on LHC energy where such a calculation is computed for the first time. The effects of the bulk pressure on thermal dilepton $v_2$ at the LHC are explored through bulk-induced modifications on the dilepton yield.
\end{abstract}


\end{frontmatter}

\section{Introduction}\label{sec:intro}

Extracting the properties of the medium created in relativistic heavy-ion experiments, be it at the Relativistic Heavy Ion Collider (RHIC) at Brook Haven National Laboratory or the Large Hadron Collider (LHC) at CERN, is at the center of modern high-energy nuclear research. The thermodynamical properties of a hot and dense strongly interacting medium at zero chemical potential, such as its speed of sound, was computed using {\it ab initio} lattice Quantum Chromodynamics ($\ell$QCD) calculations \cite{Borsanyi:2013bia,Bazavov:2014pvz}. A recent comparison, via Bayesian analysis, of Markov Chain Monte Carlo (MCMC) simulations and experimental data \cite{Pratt:2015zsa} shows that a parametrization of the speed of sound that is close to that of $\ell$QCD calculations is consistent with data, thus providing an independent method of extracting the speed of sound of QCD. In addition to its thermodynamical properties, experiments at RHIC and the LHC are also striving to extract near-equilibrium properties of QCD, specifically its dissipative transport coefficients such as bulk viscosity ($\zeta$), shear viscosity ($\eta$), net baryon number conductivity ($\kappa_B$) as well as the relaxation times associated with the bulk pressure ($\tau_\Pi$), the shear tensor ($\tau_\pi$), and net baryon number diffusion ($\tau_B$). Hadronic observables are more sensitive to some of these transport coefficients than others, for instance they are more sensitive to $\eta$ then they are to $\tau_\pi$ \cite{Song:2008si, Vujanovic:2016anq}. An increased sensitivity to all transport coefficient can be obtained via the inclusion of electromagnetic probes, such as lepton pairs, in addition to hadronic observables. Indeed, electromagnetic probes are emitted throughout the entire evolution of the medium and are therefore sensitive to earlier hotter periods where dissipation effects are larger.  

We will focus on thermal lepton pair (dilepton) production in this contribution. Thermal dileptons have shown to be sensitive to some of the above mentioned transport coefficients \cite{Vujanovic:2016anq, Vujanovic:2015nwv, Vujanovic:2014vwa}, and this study continues the efforts in that direction by investigating the influence of the bulk viscosity $\zeta$ on dilepton radiation. 

\section{Initial conditions and viscous hydrodynamics}\label{sec:visc_hydro}
The hydrodynamical simulations that are at the heart of this study use the IP-Glasma initial conditions \cite{Ryu:2015vwa,Paquet:2015lta}. The dynamical equations describing the evolution of the fluid consist of energy-momentum conservation $\partial_\mu T^{\mu\nu}=0$, and relaxation equation for the dissipative bulk pressure ($\Pi$) and shear tensor ($\pi^{\mu\nu}$) , which are given below in Eq. (\ref{eq:bulk_shear}). The energy-momentum tensor is given by $T^{\mu\nu}=\varepsilon u^\mu u^\nu - (P+\Pi)\Delta^{\mu\nu}+\pi^{\mu\nu}$, where $\varepsilon$ is the energy density, $u^\mu$ is the flow four velocity, $P$ is the thermodynamic pressure related to $\varepsilon$ by the equation of state $P(\varepsilon)$ \cite{Huovinen:2009yb}, $\Delta^{\mu\nu}=g^{\mu\nu}-u^\mu u^\nu$, and $g^{\mu\nu}={\rm diag}(+,-,-,-)$ is the metric tensor. The relaxation equations for $\Pi$ and $\pi^{\mu\nu}$ are:
\begin{eqnarray}
\tau_\Pi \dot{\Pi}+\Pi &=& -\zeta\theta \nonumber - \delta_{\Pi\Pi}\Pi\theta + \lambda_{\Pi\pi}\pi^{\alpha\beta}\sigma_{\alpha\beta},\\
\tau_\pi \dot{\pi}^{\langle\mu\nu\rangle}+\pi^{\mu\nu} &=& 2\eta\sigma^{\mu\nu}-\delta_{\pi\pi}\pi^{\mu\nu}\theta + \lambda_{\pi\Pi}\Pi\sigma^{\mu\nu}\nonumber\\
                                                          && - \tau_{\pi\pi} \pi^{\langle\mu}_\alpha\sigma^{\mu\rangle}_\alpha + \phi_7 \pi^{\langle\mu}_\alpha\pi^{\mu\rangle}_\alpha,
\label{eq:bulk_shear}
\end{eqnarray}  
where $\dot{\Pi}=u^\alpha\partial_\alpha\Pi$, $\dot{\pi}^{\langle\mu\nu\rangle}=\Delta^{\mu\nu}_{\alpha\beta}u^\lambda\partial_\lambda\pi^{\alpha\beta}$, $\Delta_{\alpha\beta}^{\mu\nu}=\left(\Delta_{\alpha}^{\mu}\Delta_{\beta}^{\nu}+\Delta_{\beta}^{\mu}\Delta_{\alpha}^{\nu}\right)/2-(\Delta_{\alpha\beta}\Delta^{\mu\nu})/3$, $\theta=\partial_\alpha u^\alpha$, and $\sigma^{\mu\nu}=\partial^{\langle\mu} u^{\nu\rangle}$. It is assumed that the ratio $\eta/s$ of shear viscosity over the entropy density ($s$) is almost temperature independent, while the bulk viscosity over entropy density ratio $\zeta/s$ is temperature-dependent as shown in Ref. \cite{Ryu:2015vwa,Paquet:2015lta}. The prescription used to fix the second-order transport coefficients is also explained in those two references. To admit a weak temperature dependence of $\eta/s$, we allow for the $\eta/s$ to change slightly between collision energies at the top RHIC energy $\sqrt{s_{NN}}=200$ GeV and at the LHC energy of $\sqrt{s_{NN}}=2.76$ TeV. The temperature dependence of specific bulk viscosity $\zeta/s$ remains the same throughout. Hydrodynamical simulations are run until the switching temperature ($T_{sw}$) between hydrodynamical degrees of freedom and hadrons is reached, following which UrQMD simulations are used to describe later dynamics. The switching temperature determined to best describe hadronic observables at top RHIC collision energy was found to be $T_{sw}=165$ MeV, whereas at LHC energy that temperature is slightly lower, namely $T_{sw}=145$ MeV \cite{Ryu:2015vwa,Paquet:2015lta}. Note that the baseline model, consisting of hydrodynamical and UrQMD simulations, is solely used when computing hadronic observables. Dilepton production, explored in the next sections, are not yet done within a hadronic transport model such as UrQMD. Hence the calculations shown herein are exploratory, as they do not yet contain the entire dynamics.    

\section{Thermal dilepton rates}\label{sec:dilepton_rates}
At leading order in the electromagnetic coupling ($\alpha$), thermal dilepton rates can be written as:
\begin{eqnarray}
\frac{d^4 R}{d^4 p} = -\frac{\alpha}{12\pi^4} \frac{1}{M^2} \frac{{\rm Im} \Pi_{{\rm R},\,\gamma^{\ast}}}{\exp\left[p\cdot u/T\right]-1},
\label{eq:rate}
\end{eqnarray}
where $T$ is the temperature of the medium, and $\Pi_{{\rm R},\,\gamma^{\ast}} = g_{\mu\nu}\Pi^{\mu\nu}_{{\rm R},\,\gamma^{\ast}}$ is the trace of the retarded virtual photon self-energy. Eq. (\ref{eq:rate}) is valid to all orders in the QCD coupling $\alpha_s$. $M^{2}=p^\mu p_\mu$, where $M$ is the virtual photon invariant mass, $p^\mu$ is the virtual photon four momentum. 

There are two sources of thermal dileptons considered in this study, one hadronic and the other partonic. The partonic dilepton emission rate is computed using the Born approximation of quark-antiquark annihilation, which is directly proportional to the quark/antiquark distribution functions \cite{Vujanovic:2013jpa}. In a dissipative medium, these distribution functions acquire viscous corrections accounting for the deformation viscosity induces on the thermal Fermi-Dirac distribution. The correction to the distribution function owing to the shear tensor is computed using the 14 moment approximation first proposed by Israel and Stewart \cite{Israel1976310,Israel:1979wp}, whereas the correction owing to bulk viscous pressure is computed using the relaxation time approximation to the modified Boltzmann equation presented in Ref. \cite{Paquet:2015lta}.  

In the hadronic medium (HM), the major source of dileptons originates from in-medium decay of vector mesons, which is well described by the Vector meson Dominance Model (VDM). VDM allows to relate the $\Pi _{\gamma^{\ast }}^{\mathrm{R}}$ to the imaginary part of the vector meson $(V)$ propagator $D_{V}^{\mathrm{R}}$. To compute the $D_{V}^{\mathrm{R}}$, the self-energy $\Pi _{V}$ of vector mesons in  the medium is required, the latter being composed of both thermal and dissipative contributions. The thermal rate and its shear viscous corrections are discussed in detail in Ref. \cite{Vujanovic:2013jpa}. Bulk viscous corrections to the thermal distribution function are computed using the relaxation time approximation of the Boltzmann equation \cite{Paquet:2015lta}. The bulk- and shear-modified thermal distribution function is in turn employed to calculate the bulk- and shear-viscous corrections to the self-energy $\Pi_V$. 

The interpolation between the hadronic and partonic degrees of freedom is done using a linear function in temperature giving the QGP fraction of a space-time cell. Consistent with the equation of state \cite{Huovinen:2009yb}, the linear interpolation is performed within the interval $184<T<220$ MeV. Dilepton emission was only evaluated above the switching temperature, which is $T_{sw}=165$ MeV at RHIC and $T_{sw}=145$ MeV at LHC energies, as mentioned in the previous section. Dilepton production from the hadronic transport dynamics will be explored and added in the near future. 

\section{Results}\label{sec:dilepton_rates}
\begin{figure}[!h]
\includegraphics[width=0.45\textwidth]{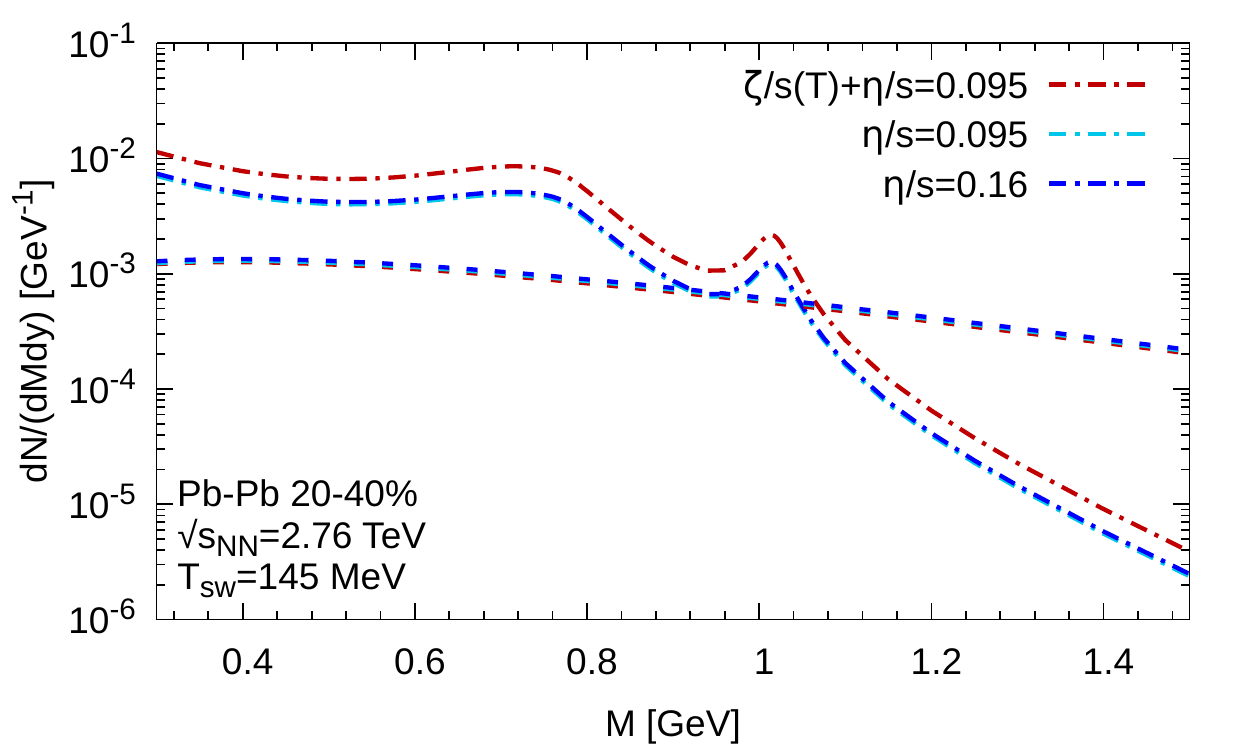}
\includegraphics[width=0.45\textwidth]{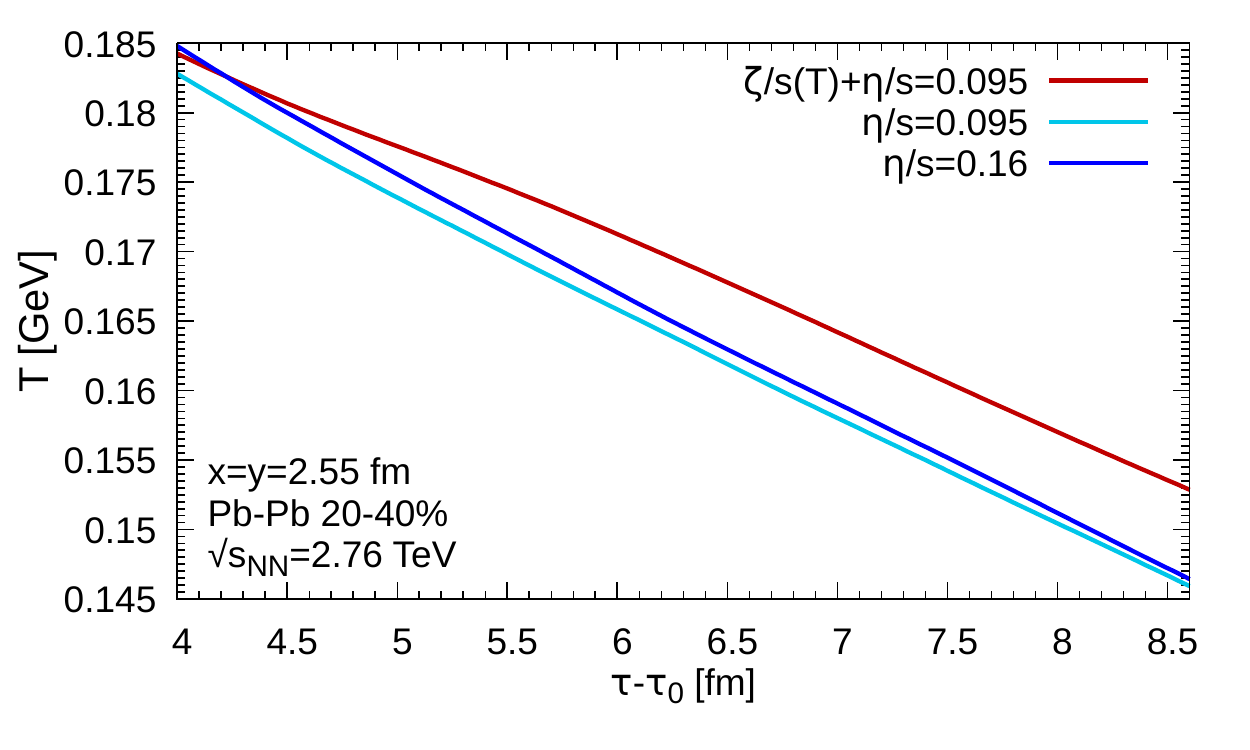}
\caption{Top panel: Dilepton yield of the HM (dash-dotted lines) and QGP (dashed lines) regions of the medium as a function of $M$ at the LHC. Bottom panel: Temperature profile for a cell at a fixed location $x=y=2.5$ fm at the LHC. For more details, see text.}
\label{fig:lhc_yield_and_Temp}
\end{figure}

The top panel of Fig. \ref{fig:lhc_yield_and_Temp} shows the dilepton radiation at LHC energies for three different media. All three simulations use the same initial condition. The medium having solely shear viscosity with $\eta/s=0.16$ (in dark blue), had its specific shear viscosity adjusted to obtain a good fit to the elliptic flow of charged hadrons \cite{Ryu:2015vwa}. The medium with bulk and shear viscosity (in red) had its specific shear viscosity lowered to $\eta/s=0.095$ in order to obtain a similarly good fit to hadronic observables. The medium in light blue uses the same shear viscosity as the one in red and its bulk viscosity was set to vanish, so it does not describe the hadronic observables as well as the other two cases. Dilepton radiation from QGP and hadronic sectors of each combination of bulk/shear viscosity are shown separately in the top panel of Fig. \ref{fig:lhc_yield_and_Temp}. 

The temperature dependence of bulk viscosity is such that it has a sharp peak at $T\sim 180$ MeV \cite{Ryu:2015vwa,Paquet:2015lta}, which occurs in the hadronic sector of the medium in our equation of state  \cite{Huovinen:2009yb}. Therefore, our $\frac{\zeta}{s}(T)$ is generating a localized increase of entropy generation, which will cause a reduced cooling rate for $T\lesssim 185$ MeV as illustrated in the bottom panel of Fig. \ref{fig:lhc_yield_and_Temp}. This leads to an increased hadronic invariant mass dilepton yield, whereas the QGP sector is barely modified. 

\begin{figure}[!h]
\includegraphics[width=0.45\textwidth]{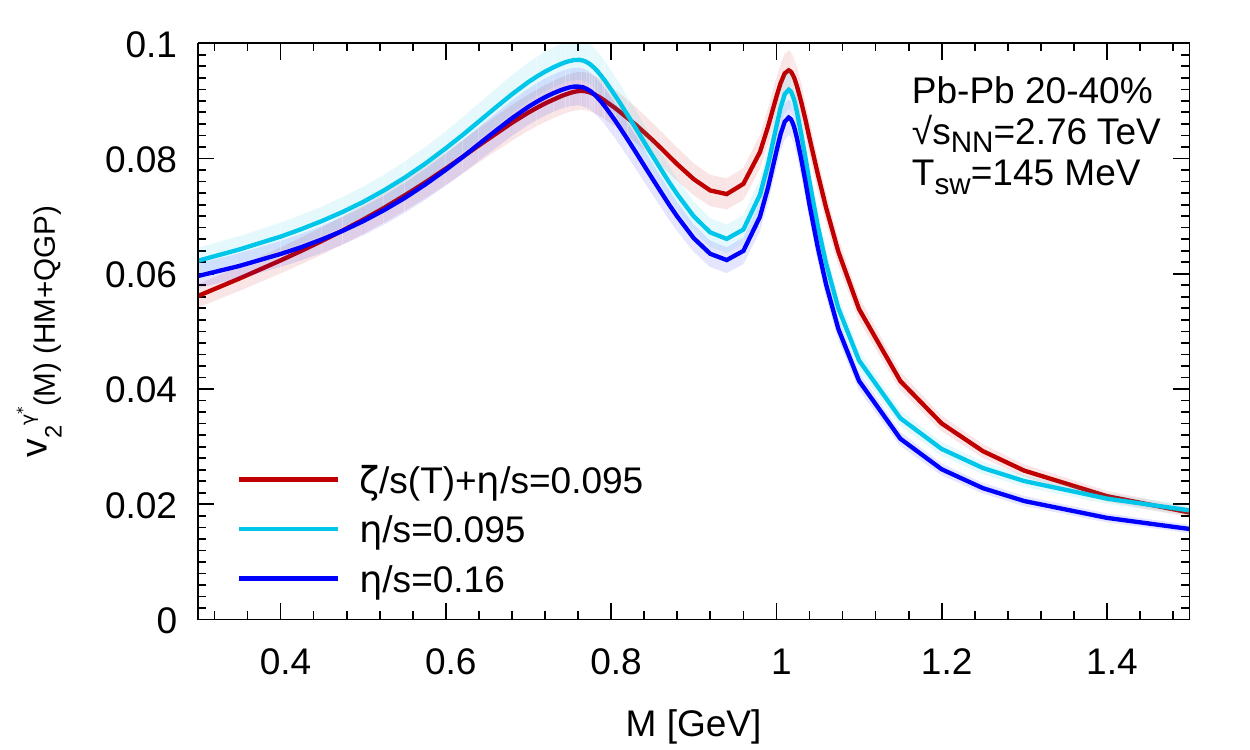}
\includegraphics[width=0.45\textwidth]{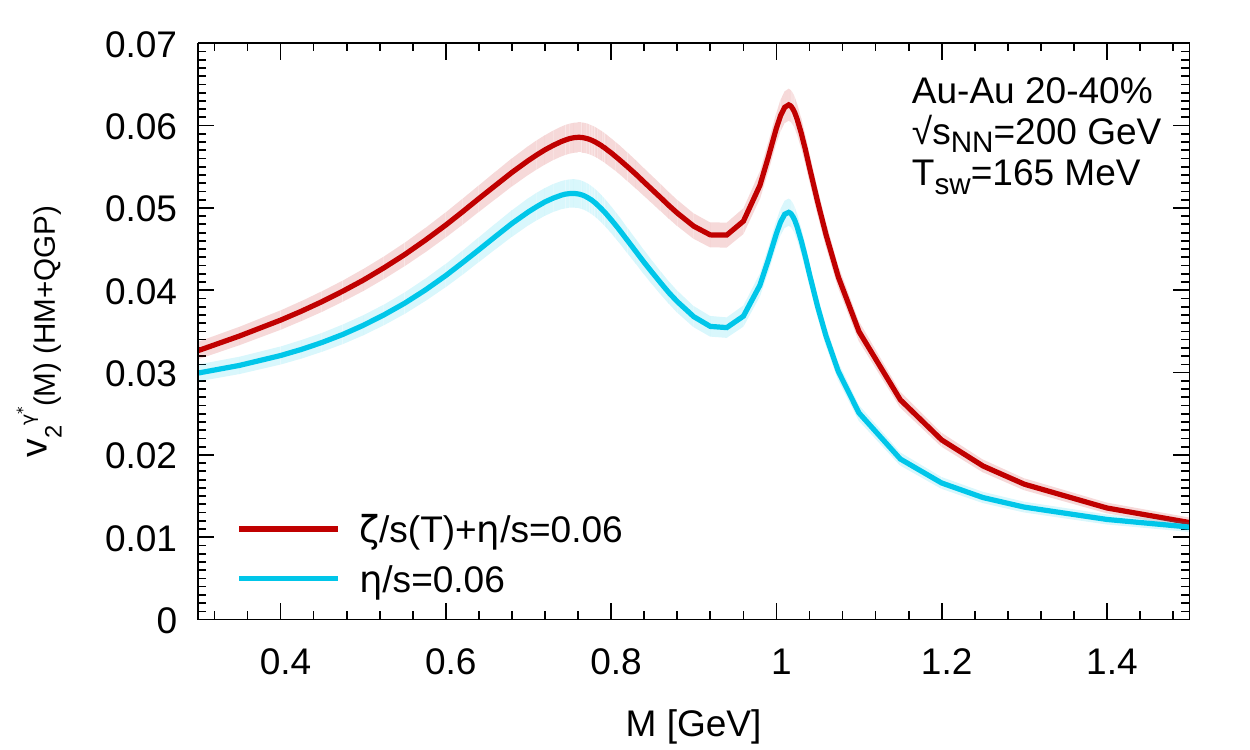}
\caption{Top panel: thermal (HM+QGP) dilepton $v_2$ as a function of $M$ at the LHC integrated for all $T>T_{sw}$. Color coding is the same as in Fig. \ref{fig:lhc_yield_and_Temp}. Bottom panel: thermal (HM+QGP) dilepton $v_2$ at RHIC integrated for all $T>T_{sw}$, where both curves use the same initial condition.}
\label{fig:v2_M}
\end{figure}

The total thermal  (partonic and hadronic) dilepton $v_2$ is shown in the top panel of Fig. \ref{fig:v2_M}. The interesting features seen in the thermal $v_2(M)$ for $M>0.85$ GeV are driven by the modification of the yield of hadronic medium (HM) versus QGP dileptons induced by the presence of bulk viscosity. It is in the region $M>0.85$ GeV that the yield goes from being HM dominated to having a significant QGP contribution. Therefore, the fact that the HM dilepton yield for the media with shear and bulk viscosity is greater than the HM dilepton yield with just shear viscosity becomes important (see top panel of Fig. \ref{fig:lhc_yield_and_Temp}). Indeed, bulk viscosity reduces the expansion rate during the hadronic stage of the evolution, giving more weight (through a larger space-time volume) to the stronger elliptic flow in the hadronic stage, thereby increasing the total $v_2$ for $M>0.85$ GeV.    

At top RHIC energy, the value of $\eta/s$ needed to be reduced to 0.06 in order to reproduce hadronic observables \cite{Ryu:2015vwa,Paquet:2015lta}. Using this value of $\eta/s$, the dilepton radiation at the RHIC is shown in the bottom panel of Fig. \ref{fig:v2_M}. The fact that thermal dilepton $v_2$ of the medium with bulk and shear viscosity increases for all $M$ relative to the medium with just shear viscosity, has to be related with the fact that $T_{sw}$ at RHIC is higher than at LHC. At RHIC, dileptons become more sensitive to the dynamical evolution of the medium occurring at higher temperatures, owing to the larger $T_{sw}$, which reduces the hadronic contribution. At LHC energy, the weight of the high- relative to low-temperature dynamics on the total dilepton yield is reduced, since $T_{sw}$ is lowered and thus the volume of HM dilepton radiation is dynamically increased. So, the size of the bulk-induced modification to the development of anisotropic flow is related to the interplay between $T_{sw}$ and the sharp peak in temperature of $\zeta/s$. An in-depth exploration of these novel dynamics is beyond the scope of these proceedings and will be done in an upcoming publication. 

To determine whether experimental observations at RHIC will be sensitive to the bulk viscosity effects seen in our $v_2$ calculation, a more complete calculation including dilepton emission from the low density post-hydrodynamical hadronic medium, which is presently neglected, should be added. To this end, a dilepton calculation from the hadronic transport model is currently being done.  

\section{Conclusion}\label{sec:conc}
In this contribution, we have shown that thermal dileptons are highly sensitive to the details of the hydrodynamical evolution both at RHIC and at LHC energies. While the behavior of dilepton elliptic flow at the LHC is driven by the increase of the yield of the HM, the behavior of $v_2$ at top RHIC energy needs to be investigated further. Whether or not this behavior can be measured at RHIC depends on how much dilepton yield and anisotropic flow is generated during hadronic transport phase of the medium, which will be explored by including a dilepton calculation from hadronic transport in addition to the current hydrodynamical simulation.

\section*{Acknowledgements}
This work was supported in part by the Natural Sciences and Engineering Research Council of Canada, by the Director, Office of Energy Research, Office of High Energy and Nuclear Physics, Division of Nuclear Physics, of the U.S. Department of Energy under awards No. \rm{DE-SC0004286}, \rm{DE-FG02-88ER40388}, and by the National Science Foundation (in the framework of the JETSCAPE Collaboration) through award No. 1550233. G. V., G. S. D., and  C. G. gratefully acknowledge support by the Fonds de Recherche du Qu\'ebec --- Nature et les Technologies (FRQNT), through the Banting Fellowship from the Government of Canada, and from the Canada Council for the Arts through its Killam Research Fellowship program, respectively. Computations were performed on the Guillimin supercomputer at McGill University under the auspices of Calcul Qu\'ebec and Compute Canada. The operation of Guillimin is funded by the Canada Foundation for Innovation (CFI), the National Science and Engineering Research Council (NSERC), NanoQu\'ebec, and the Fonds de Recherche du Qu\'ebec --- Nature et les Technologies (FRQNT).

\nocite{*}
\bibliographystyle{elsarticle-num}
\bibliography{references_short}

\begin{thebibliography}{10}
\expandafter\ifx\csname url\endcsname\relax
  \def\url#1{\texttt{#1}}\fi
\expandafter\ifx\csname urlprefix\endcsname\relax\def\urlprefix{URL }\fi
\expandafter\ifx\csname href\endcsname\relax
  \def\href#1#2{#2} \def\path#1{#1}\fi

\bibitem{Borsanyi:2013bia}
S.~Borsanyi, Z.~Fodor, C.~Hoelbling, S.~D. Katz, S.~Krieg, K.~K. Szabo, Phys.
  Lett. B730 (2014) 99--104.
\newblock \href {http://arxiv.org/abs/hep-lat/1309.5258}
  {\path{arXiv:hep-lat/1309.5258}}.

\bibitem{Bazavov:2014pvz}
A.~Bazavov, et~al., Phys. Rev. D90 (2014) 094503.
\newblock \href {http://arxiv.org/abs/hep-lat/1407.6387}
  {\path{arXiv:hep-lat/1407.6387}}.

\bibitem{Pratt:2015zsa}
S.~Pratt, E.~Sangaline, P.~Sorensen, H.~Wang, Phys. Rev. Lett. 114 (2015)
  202301.
\newblock \href {http://arxiv.org/abs/nucl-th/1501.04042}
  {\path{arXiv:nucl-th/1501.04042}}.

\bibitem{Song:2008si}
H.~Song, U.~Heinz, Phys. Rev. C78 (2008) 024902.
\newblock \href {http://arxiv.org/abs/nucl-th/0805.1756}
  {\path{arXiv:nucl-th/0805.1756}}.

\bibitem{Vujanovic:2016anq}
G.~Vujanovic, J.-F. Paquet, G.~S. Denicol, M.~Luzum, S.~Jeon, C.~Gale, Phys.
  Rev. C94~(1) (2016) 014904.
\newblock \href {http://arxiv.org/abs/nucl-th/1602.01455}
  {\path{arXiv:nucl-th/1602.01455}}.

\bibitem{Vujanovic:2015nwv}
G.~Vujanovic, C.~Shen, G.~S. Denicol, B.~Schenke, S.~Jeon, C.~Gale, in:
  {Proceedings, 7th International Conference on Hard and Electromagnetic Probes
  of High-Energy Nuclear Collisions (Hard Probes 2015): Montréal, Québec,
  Canada, June 29-July 3, 2015}, 2016.
\newblock \href {http://arxiv.org/abs/nucl-th/1511.04625}
  {\path{arXiv:nucl-th/1511.04625}}.

\bibitem{Vujanovic:2014vwa}
G.~Vujanovic, J.-F. Paquet, G.~S. Denicol, M.~Luzum, B.~Schenke, S.~Jeon,
  C.~Gale, Nucl. Phys. A931 (2014) 701--705.
\newblock \href {http://arxiv.org/abs/nucl-th/1408.1098}
  {\path{arXiv:nucl-th/1408.1098}}.

\bibitem{Ryu:2015vwa}
S.~Ryu, J.~F. Paquet, C.~Shen, G.~S. Denicol, B.~Schenke, S.~Jeon, C.~Gale,
  Phys. Rev. Lett. 115~(13) (2015) 132301.
\newblock \href {http://arxiv.org/abs/nucl-th/1502.01675}
  {\path{arXiv:nucl-th/1502.01675}}.

\bibitem{Paquet:2015lta}
J.-F. Paquet, C.~Shen, G.~S. Denicol, M.~Luzum, B.~Schenke, S.~Jeon, C.~Gale,
  Phys. Rev. C93~(4) (2016) 044906.
\newblock \href {http://arxiv.org/abs/hep-ph/1509.06738}
  {\path{arXiv:hep-ph/1509.06738}}.

\bibitem{Huovinen:2009yb}
P.~Huovinen, P.~Petreczky, Nucl. Phys. A837 (2010) 26--53, hep-ph/0912.2541.

\bibitem{Vujanovic:2013jpa}
G.~Vujanovic, C.~Young, B.~Schenke, R.~Rapp, S.~Jeon, et~al., Phys. Rev.
  C89~(3) (2014) 034904.
\newblock \href {http://arxiv.org/abs/nucl-th/1312.0676}
  {\path{arXiv:nucl-th/1312.0676}}.

\bibitem{Israel1976310}
W.~Israel, Annals of Physics 100~(1–2) (1976) 310 -- 331.
\newblock \href
  {http://dx.doi.org/http://dx.doi.org/10.1016/0003-4916(76)90064-6}
  {\path{doi:http://dx.doi.org/10.1016/0003-4916(76)90064-6}}.

\bibitem{Israel:1979wp}
W.~Israel, J.~Stewart, Annals Phys. 118 (1979) 341--372.
\newblock \href {http://dx.doi.org/10.1016/0003-4916(79)90130-1}
  {\path{doi:10.1016/0003-4916(79)90130-1}}.

\end{thebibliography}
\end{document}